\documentstyle[osa,psfig]{revtex}

\begin{document}

\title{Analysis of astronomical data from optical superconducting tunnel junctions}

\author{J.H.J.\ de Bruijne, A.P.\ Reynolds, M.A.C.\ Perryman, F.\ Favata, A.\ Peacock}

\address{Astrophysics Division, Space Science Department of ESA,
ESTEC,
P.O.Box 299,
NL--2200 AG Noordwijk,
the Netherlands;
jdbruijn@astro.estec.esa.nl
{\rm{\centerline{\null}}}\hfill\break
{\rm{\centerline{Accepted for publication in: {\sl Focal plane detector array developments}, eds Z.\ Ninkov, W.J.\ Forrest,}}}\hfill\break
{\rm{\centerline{Optical Engineering (The International Society for Optical Engineering; SPIE),}}}\hfill\break
{\rm{\centerline{scheduled for publication in December 2001.}}}\hfill\break
}

\maketitle

\begin{abstract}

\noindent Currently operating optical superconducting tunnel junction
(STJ) detectors, developed in ESA, can simultaneously measure the
wavelength ($\Delta\lambda = 50$~nm at $500$~nm) and arrival time (to
within $\sim$$5~\mu$s) of individual photons in the range 310--720~nm
with an efficiency of $\sim$70\%, and with count rates of order
$5\,000$~photons per second per junction. A number of STJ junctions
placed in an array format generates four-dimensional data: photon
arrival time, energy, and array element $(X,Y)$. Such STJ cameras are
ideally suited for, e.g., high time-resolution spectrally-resolved
monitoring of variable sources or low-resolution spectroscopy of faint
extragalactic objects.

The reduction of STJ data involves detector efficiency correction,
atmospheric extinction correction, sky background subtraction, and,
unlike that of data from CCD-based systems, a more complex energy
calibration, barycentric arrival time correction, energy range
selection, and time binning; these steps are, in many respects,
analogous to procedures followed in high-energy astrophysics. This
paper discusses these calibration steps in detail using a
representative observation of the cataclysmic variable UZ Fornacis;
these data were obtained with ESA's S--Cam2 $6 \times 6$-pixel
device. We furthermore discuss issues related to telescope pointing
and guiding, differential atmospheric refraction, and
atmosphere-induced image motion and image smearing (``seeing'') in the
focal plane. We also present a simple and effective recipe for
extracting the evolution of atmospheric seeing with time from any
science exposure, and discuss a number of caveats in the
interpretation of STJ-based time-binned data, such as light curves and
hardness ratio plots.


\end{abstract}

\section{Introduction}
\label{sec:1}

Superconducting tunnel junctions (STJs) are powerful photon counting
detectors. Since the energy gap between the ground state and the first
excited state of an STJ is only a few meV (compared to a band gap of
$\sim$1~eV for a semi-conductor such as silicon), an individual
optical photon, with an energy of order a few eV, has the ability to
free a large number of charge carriers, in proportion to its energy,
when hitting an STJ. Measuring the resulting electrical pulse allows
for an accurate and simultaneous determination of the photon arrival
time and energy.

A number of STJs placed in an array format provides a superconducting
tunnel junction camera (S--Cam). Such an instrument is currently being
developed within the Astrophysics Division of the Space Science
Department of ESA (e.g., Peacock et al.\ 1997, 1998; Verhoeve et al.\
2001, in this volume). The presently operating prototype, S--Cam2, has
seen a successful series of technical validation and science
qualification runs at the 4.2~m William Herschel Telescope (WHT) on La
Palma.

S--Cam2 consists of a staggered $6 \times 6$ array of Tantalum STJs
measuring $25 \times 25~\mu$m$^2$ each (Figure~\ref{fig:1}).
Inter-junction dead spaces measure $\sim$$4~\mu$m, resulting in a
filling factor of $\sim$0.78. In the focal plane, each junction
corresponds to $\sim$$0.6^{\prime\prime} \times 0.6^{\prime\prime}$ on
the sky; the entire field of view of S--Cam2 is
$\sim$$4^{\prime\prime} \times 4^{\prime\prime}$. The focal plane
array is directly illuminated from the f/11 Nasmyth~1 focus of the
telescope, without loss of angular resolution, through dedicated
collimating optics and a lateral optical window. The array operates at
$\sim$0.32~K, well below the critical temperature of Tantalum (4.48~K)
in a sophisticated cryogenic environment; instrumental dark current,
the amplitude of which depends on the detector temperature and on any
residual infrared stray radiation, is negligible (Verhoeve et al.\
2001). The photon-generated electrical pulses are read out using
charge-sensitive amplifiers operating at room temperature; S--Cam2,
being a photon-counting detector, does not suffer from read-out
noise. A dedicated data acquisition computer stores the observations,
in real time, in a 4-column event-list format: photon arrival time,
energy, and array element $(X,Y)$.

Whereas the intrinsic speed of a Tantalum STJ is typically below
$\sim$$1$~ns, S--Cam2 photon arrival time stamps are accurate to
within $\sim$$5~\mu$s with respect to GPS timing signals; this
accuracy level is set by the present signal processing
electronics. S--Cam2 has a count rate limitation of
$\sim$$5\,000$~photons~s$^{-1}$ per junction and
$\sim$$30\,000$~photons~s$^{-1}$ for the entire device; these limits
are also set by the present electronics and real-time processing
software. The spectral resolving power $\lambda / \Delta \lambda$ of
an STJ increases with the square root of the photon energy. The
intrinsic wavelength resolution, the so-called Fano limit, of Tantalum
is $\Delta \lambda \sim 12$~nm at $\lambda = 500$~nm. The
corresponding value for S--Cam2 is $\sim$60~nm at 500~nm; this
value is mainly set by two, roughly equally important, components: the
electronics and the thermal infrared background radiation from the
telescope and atmosphere. Although the intrinsic wavelength response
of Tantalum is very broad (from shortward of 300~nm to longward of
$2\,000$~nm), the response of S--Cam2 is restricted to
$\sim$310--720~nm. The blue limit is predominantly set by the Earth's
atmosphere. In order to reduce thermal noise photons, all low-energy
(i.e., red) photons are deliberately suppressed by means of a series
of optical filter elements.

The photon detection efficiency (``quantum efficiency'') of S--Cam2 is
$\sim$60--70\% over the range 310--720~nm; it is mainly determined by
the properties of the substrate (sapphire) on which the
back-illuminated STJs are deposited. The total throughput of the
S--Cam2 optical system, including the WHT primary and secondary
mirrors ($\sim$$85$\% transmission each), the Nasmyth flat
($\sim$$85$\% transmission) and optical derotator ($\sim$$50$\%
transmission), the S--Cam2 optics ($\sim$$25$\% transmission), and the
detection efficiency ($\sim$$70$\%), is a few per cent over this
range.

The high detection efficiency, broad energy response, and high speed
of STJ cameras, combined with their inherent energy resolution and
photon counting capabilities, make these instruments ideally suited to
perform, e.g., high time-resolution energy-resolved photometry of
variable sources (e.g., Perryman et al.\ 2001) or low-resolution
spectroscopy of faint objects (``photometric redshifts''; e.g.,
Perryman et al.\ 1993; Jakobsen 1999; Mazin et al.\ 2000). S--Cam2
observations of a variety of astronomical objects, such as
$\gamma$-ray bursts, cataclysmic variables, pulsars, and dwarf novae,
are currently being analyzed in detail. Future investigations will use
S--Cam data to deepen the study of the characteristics of atmospheric
seeing and scintillation, touching upon topics such as speckle imaging
and spectroscopy, adaptive optics, and interferometric fringe
detection.

STJ cameras are complex systems, both from the manufacturing and the
operational point of view (see, e.g., Verhoeve et al.\ 2001 for a
detailed discussion). Moreover, the four-dimensional data cube
generated by STJ detectors provides new challenges regarding data
reduction and analysis as compared to normal CCD procedures, although
general experience and procedures built up in and developed for the
field of high-energy astrophysics have proven highly useful. This
contribution addresses some of these aspects by using a representative
S--Cam2 observation of the cataclysmic variable UZ For as a working
example. Astrophysical results of this and other S--Cam2 exposures
will be presented elsewhere.

\section{Reduction and calibration}
\label{sec:2}

S--Cam data are stored, in conventional binary FITS format, as event
lists, containing for each detected photon a spatial location (array
element $X,Y$ in the range $1$--$6,1$--$6$), arrival time, and energy
channel (in the range 0--255). Prior to scientific analyses, raw
S--Cam data are fed through an automated pipeline. This software
performs a pixel-dependent energy calibration, a photon time stamp
correction (when referring to the Solar system barycentre), a
pixel-dependent detection efficiency correction (``flat fielding''),
an energy-dependent atmospheric extinction correction, and a sky
background subtraction. The following subsections summarize each of
these steps, using an exposure representative of the characteristics
and capabilities of S--Cam2 as an example. This observation, of the
eclipsing binary system UZ Fornacis consisting of a white dwarf and a
main sequence star, was taken on October 4, 2000, 04:30:41--05:00:41
UTC under moderate seeing conditions ($\sim$$1^{\prime\prime}$; \S
\ref{sec:4F}). During the exposure, UZ For's zenith angle evolved from
$z = 55.19^{\circ}\!$ to $56.78^{\circ}\!$, i.e., the airmass $A$
changed from $1.75$ to $1.83$. The $1\,800$-s frame captures, at
roughly mid-exposure, a $\sim$$480$~s eclipse of the bright white
dwarf. During this event, the relatively faint reddish contribution
from the low-mass companion star remains visible (\S \ref{sec:4B};
cf.\ Reynolds 2001).

\subsection{Energy calibration}
\label{sec:2A}

Incident photons of energy $E_{\rm photon}$ are assigned to energy
channels $E_i = i$, where $i = 0, \ldots, 255$. Laboratory
measurements using a tunable monochromatic light source (325--725~nm)
confirm that all 36 S--Cam2 junctions have a highly linear, yet
slightly pixel-dependent, energy response:
\begin{equation}
E_i^j = G^j \cdot E_{\rm photon} + C^j,
\label{eq:1}
\end{equation}
where $G^j$ (in channels per eV) is the gain and $C^j$ (in channels)
the corresponding offset of pixel $j$. We have implemented the energy
calibration of S--Cam2 in two steps. First we correct for the
inter-pixel differences by bringing the gain and offset of all pixels
to that of an arbitrary reference pixel ($X,Y = 3,1$ in this case). We
then assign an absolute energy (or wavelength) calibration to this
reference pixel.

As the monochromator is a non-portable device, the exact values and
temporal stability of the different gains and offsets during an
observing campaign can only be obtained and monitored by using a red
LED as a semi-monochromatic secondary light source ($\lambda = 650$~nm
at room temperature). When operating this LED at a sufficiently high
photon flux, two (or~$n$) equal-energy LED photons arriving
simultaneously, i.e., within a few $\mu$s of each other, are detected
as one event with twice (or $n$ times) the single LED photon
energy. An LED spectrum thus consists of a primary peak, corresponding
to 650~nm photons, and a number of higher-order photon peaks at
multiple energies of a single LED photon. In practice, only two
higher-order peaks are detected in all pixels, of which only the first
one is strong enough to be used. By measuring the centroids of these
two peaks, $E_{\rm peak~1}^j$ and $E_{\rm peak~2}^j$, for each pixel
$j$, the observed energies $E_i^j$ of all detected photons can be
brought to a common reference scale, corresponding to pixel $(X,Y =
3,1)$, according to:
\begin{equation}
E_i^j \rightarrow {\rm NINT}\left(E_{\rm peak~1}^{(3,1)} + {{E_{\rm peak~1}^{(3,1)} - E_{\rm peak~2}^{(3,1)}}\over{E_{\rm peak~1}^{j} - E_{\rm peak~2}^{j}}} \cdot \left(E_i^j - 0.5 + {\rm RAN} - E_{\rm peak~1}^{(3,1)}\right)\right),
\label{eq:2}
\end{equation}
where RAN is a random deviate between 0 and 1, and the function NINT
rounds to the nearest integer. The term ``$-0.5 + {\rm RAN}$'', which
is a random number between $-0.5$ and $+0.5$, has been included,
following normal practice in high-energy astrophysics analyses (e.g.,
Hasinger \& Snowden 1990), in order to avoid a repetitive pattern of
``spikes'' and/or ``dips'' in the gain-corrected energy channel
distributions (the energy resolution decrease resulting from this step
is negligible).

The absolute calibration of the reference pixel can be established by
using the laboratory-based monochromator data as well as the LED
exposures taken throughout the observing campaign (the calibration is
somewhat complicated in practice as a result of the presence of a
small but significant dependence of the LED photon energy with ambient
temperature). These methods provide consistent results, $G^{(3,1)}
\sim 42.5$ and $C^{(3,1)} \sim -2.0$, confirming the high temporal
stability of the detector. The final (pixel-independent) relation
between (gain-corrected) energy channel $E_i = i~ (i = 0, \ldots,
255)$ and photon wavelength $\lambda$ thus becomes:
\begin{equation}
\lambda~[{\rm nm}] = {{1\,239.75 \cdot G^{(3,1)}}\over{E_i - C^{(3,1)}}} = {{52\,689.4}\over{E_i + 2.0}},
\label{eq:3}
\end{equation}
where $1\,239.75$ is the conversion factor between photon wavelength
$\lambda~[{\rm nm}]$ and energy $E~ [{\rm eV}]$.

\subsection{Barycentric time stamp correction}
\label{sec:2B}

Photon arrival times are stored in UTC (Universal Coordinated Time).
However, studies of time-dependent targets such as binary systems or
pulsars often require arrival times with respect to the Solar system
barycentre (TDB, Barycentric Dynamical Time). The pipeline processing
therefore optionally modifies UTC time stamps to TDB values. It does
so by going from UTC to TAI (International Atomic Time; $+32$~s at the
observation epoch of UZ For), from TAI to TDT (Terrestrial Dynamical
Time; $+32.184\,000$~s), and from TDT to TDB. The last step requires
knowledge of the right ascension and declination of the object at the
observation epoch and the geographic longitude, latitude, and altitude
of the telescope to correct for the observer's location on and the
diurnal motion of the Earth. Our software uses the JPL DE200 ephemeris
to track the location of the Earth with respect to the Solar system
barycentre during the observation, taking propagation delay in
space--time, caused by the presence of the Sun, into account. For UZ
For, the final corrections applied to the arrival time stamps vary
from $+369.799\,990$~s for the first to $+369.865\,824$~s for the last
photon.

\subsection{Energy range selection}
\label{sec:2C}

In the next step, the photon stream is split into a number of
energy-selected ranges. This is useful for rapid diagnostic purposes
and allows for the construction of, e.g., colour-colour and hardness
ratio plots (\S \ref{sec:4B}). The pipeline software generates, on the
basis of the shape of the underlying source spectrum, a user-defined
number of energy bands, so that (roughly) equal numbers of events are
contained in each of them. The default number of ranges is presently
$n = 3$ (given the limited energy resolution of S--Cam2), which are
loosely referred to here as ``red'', ``yellow'', and ``blue''. For the
case of UZ For, the low-energy band is $E_{0}$--$E_{98}$ (red), the
medium-energy band is $E_{99}$--$E_{116}$ (yellow), and the
high-energy band is $E_{117}$--$E_{255}$ (blue; cf.\
Table~\ref{tab:1}). From now on, all calibration steps are performed
on the data file containing all counts (energy band 0) as well as on
the $n$ energy-selected sub-exposures.

\subsection{Time binning}
\label{sec:2D}

The remaining calibration steps (detection efficiency correction,
atmospheric extinction correction, and sky background subtraction) are
intricate to introduce and deal with at the individual photon level.
The pipeline software therefore introduces an {\it a posteriori\/}
time binning of the data, based on a user-specified time
interval. This interval can, in principle, be arbitrarily small or
large, and should be defined taking into account the specific
scientific question which is to be addressed. For the UZ For data
discussed here, we adopt a $1$-s binning. Figure~\ref{fig:1} shows the
resulting pixel-selected light curves at this stage (cf.\ \S
\ref{sec:4B}).

\subsection{Detection efficiency correction (``flat fielding'')}
\label{sec:2E}

The detection efficiency of an S--Cam pixel is, in principle, a
function of time, the incident-photon energy, and the pixel
number. The latter dependency results from intrinsic pixel-to-pixel
differences as well as vignetting effects related to the relay optics
including the infrared photon blocking filters. Laboratory
measurements have shown that detection efficiency variations are
largely time- and energy-independent, so that, in practice, the usage
of a single ``flat field map'' is adequate.

We therefore choose, in order to correct for pixel-to-pixel efficiency
variations, to apply a differential correction, relative to an
arbitrary reference pixel ($X,Y = 3,1$ in this case), to the
time-binned count rates (``exposure time correction'' in high-energy
astrophysics jargon). Figure~\ref{fig:2} shows a correction map which
is based on 18 sky observations obtained during the third S--Cam2
campaign (Sep--Oct 2000). Its smooth structure is most likely
explained in terms of vignetting, suggesting that intrinsic
pixel-to-pixel variations are probably small.

\subsection{Atmospheric extinction correction}
\label{sec:2F}

In the optical part of the spectrum, atmospheric extinction is a
function of wavelength as a result of dust scattering and of Rayleigh
scattering and absorption by air molecules (e.g., Hayes et al.\
1975). Furthermore, extinction is zenith angle- and thus pointing- and
time-dependent. It is therefore important to correct for this
effect. Atmospheric extinction affects the ground-based observed
magnitude $V_{\lambda}$ ``at wavelength $\lambda$'' of a celestial
object according to:
\begin{equation}
V_{\lambda} = V_{0, \lambda} + k_{\lambda} \cdot A,
\label{eq:4}
\end{equation}
where $V_{0, \lambda}$ is the object's magnitude above the Earth's
atmosphere, $k_{\lambda} = 1.086 \cdot \tau_0$~[mag~airmass$^{-1}$] is
the extinction coefficient, $\tau_0 > 0$ is the optical depth of the
atmosphere at zenith angle $z = 0^{\circ}\!$ ($A = 1$), and $A \sim
{\rm sec}(z) = {\cos}^{-1}(z)$ is the airmass\footnote{
This formula is based on the plane parallel layer approximation for
the atmosphere. The pipeline software uses an extended formula, which
is also valid at large zenith angles ($z \mathrel{{\hbox to 0pt{\lower
3pt\hbox{$\sim$}\hss}}\raise2.0pt\hbox{$>$}} 80^{\circ}\!$): $A = {\rm
sec}(z) - 0.001\,816\,7 \cdot ({\rm sec}(z)-1) - 0.002\,875\,0 \cdot ({\rm
sec}(z)-1)^2 - 0.000\,808\,3 \cdot ({\rm sec}(z)-1)^3$.
}.

It is customary, partly because more rigorous options are impractical,
to correct for extinction by using time-averaged extinction
coefficients. Although extinction can vary from night to night and
from season to season, mainly due to a varying contribution from dust
scattering (but see Stickland et al.\ 1987), the pipeline software
uses mean coefficients $k_\lambda$ ($\lambda = 300$--$1100$~nm) valid
for La Palma (King 1985). For each energy band, the wavelength
corresponding to the centroid energy channel is calculated according
to equation~3, after which $k_\lambda$ is determined using a linear
interpolation in the La Palma table. The time-binned count rates are
subsequently multiplied by $10^{0.4 k_{\lambda} \cdot
A}$. Table~\ref{tab:1} summarizes this calibration step for the UZ For
exposure discussed here.

\subsection{Background subtraction}
\label{sec:2G}

In the optical part of the spectrum, the main contributions to the
intensity of the night sky are, in order of decreasing importance, the
Moon (e.g., Krisciunas et al.\ 1991), airglow, and zodiacal light
(e.g., Benn et al.\ 1998; Leinert et al.\ 1998). As a result, the
night sky brightness is dependent on pointing, and, as a consequence,
on time\footnote{
Intrinsic variability is driven by, e.g., the amount of dust in the
atmosphere and the phase of the Solar activity cycle.
}. The time variability is, however, generally negligibly small when
tracking an object on time scales of tens of minutes, except when
observing during a Moon-rise or -set\footnote{
A time-variable background intensity can also result from
telescope-induced guiding motions when observing, e.g., objects
superimposed on extended extragalactic sources, or objects embedded in
a spatially non-uniform nebulosity (e.g., the Crab pulsar).
}.

An optimum background subtraction, i.e., source extraction, for S--Cam
data requires the construction of a ``spatial source mask''. This is a
non-trivial exercise, as atmospheric seeing affects both the size and
the location of the point spread function on time scales of ms (e.g.,
Coulman 1985; cf.\ \S \ref{sec:4F}). The location of such a source
mask is, moreover, energy-dependent as a result of differential
atmospheric refraction (\S \ref{sec:4D}). As a first step, we have
therefore implemented a simplified background subtraction procedure.

The S--Cam data reduction software, by default, applies a
time-independent background subtraction, and optionally allows for a
time-dependent sky brightness determination. Under normal atmospheric
seeing, telescope guiding, and image centering conditions, the corner
pixels $(X,Y = 1,1)$ and $(6,6)$ are, as a result of the S--Cam2
device architecture, most suitable for a determination of the sky
background intensity (e.g., Figure~\ref{fig:1}). The bottom panels of
Figure~\ref{fig:3} show, for the Moon-less UZ For exposure, the light
curves of these two pixels at the selected binning period (1~s; \S
\ref{sec:4B}). These junctions clearly provide a robust estimate of
the sky brightness. The default procedure therefore is to subtract,
from the pixel-selected light curves, the mean value of the
time-averaged count rates of the two corner pixels
($\sim$$26.5$~counts~s$^{-1}$~pixel$^{-1}$ in this case); the
time-dependent option subtracts the mean value of the time-binned
count rates, although subtracting a smoothly varying function fitted
to the time-binned count rates has the advantage that the
signal-to-noise ratio of the resulting light curve is not degraded
unnecessarily. The top panel of Figure~\ref{fig:3} shows the
pixel-integrated sky-subtracted light curve of UZ For (all energies;
cf.\ Figure~\ref{fig:4}).

\section{Light curve interpretation}
\label{sec:3}

The light curves of UZ For displayed, e.g., in Figure~\ref{fig:3} show
a rich structure. Before attributing this structure to physical
properties of the observed source, it is important to address a number
of potential caveats in the interpretation of STJ-based time-binned
data. Whereas some of these caveats are well known and common to the
general field of aperture photometry (e.g., Mighell 1999), others are
specific to S--Cam observations.

\subsection{Photon noise and atmospheric scintillation}
\label{sec:3A}

A fundamental limit to the precision with which the brightness of any
celestial object can be measured is set by Poissonian photon
noise. The ``flat-fielded'' and sky-subtracted light curve of UZ For
(e.g., Figure~\ref{fig:3}) has, outside eclipse, a count rate of
$\sim$$2\,000$~counts per 1-s time bin. The correspondingly expected
Poisson noise is $\sqrt{2\,000} = 45$~counts~s$^{-1}$, i.e.,
$\sim$$2.2$\%.

Atmospheric scintillation (the rapidly varying turbulent refocussing
of light rays passing through the atmosphere) is a second random noise
contributor beyond the control of the observer. For integration times
of order $\sim$$1$~s or longer, the root-mean-square deviation of the
relative intensity of star light due to (the low-frequency component
of) scintillation is given by (e.g., Young 1967; Young et al.\ 1991;
Dravins et al.\ 1998):
\begin{equation}
\sim\ 0.09~ \left({{D}\over{1~{\rm cm}}}\right)^{-2/3}~ {\rm cos}^{-7/4}(z)~ \exp\left(-{{L}\over{L_0}}\right)~ \left({{2~T}\over{1~{\rm s}}}\right)^{-1/2} \sim\ 0.2\%,
\label{eq:5}
\end{equation}
where $D = 4.2$~m, $z \sim 56^{\circ}\!$, $L_0 = 8\,000$~m is the
pressure scale height of the atmosphere, $L = 2\,332$~m is the
altitude of the aperture above sea level, and $T = 1$~s is the
integration time (of a single time bin in this case). We thus conclude
that Poisson noise is the dominant random noise term. This is
generally true for a 1-s time binning: scintillation starts dominating
Poisson noise at the WHT when receiving more than
$\sim$$1\,391\,000/123\,000$~photons~s$^{-1}$ from a source at zenith
angle $z = 0^{\circ}\!/60^{\circ}\!$ (recall that S--Cam2 can ``only''
handle count rates up to $\sim$$30\,000$~photons~s$^{-1}$ due to
electronic limitations).

\subsection{Caveats related to image motion}
\label{sec:3B}

Atmospheric seeing is known to affect both the size and the location
of the point spread function at time scales of ms (e.g., Coulman 1985;
cf.\ \S \ref{sec:4F}). This is clearly demonstrated by the
pixel-selected light curves displayed in Figure~\ref{fig:1}. Image
motion (\S \ref{sec:4E}), either due to atmospheric seeing or due to
guiding and tracking errors and corrections (cf.\ \S \ref{sec:4G}),
may lead to spurious effects in, e.g., (energy-selected) light curves
and hardness ratio plots (\S \ref{sec:4B}) for at least three reasons.

First of all, possibly uncorrected spatial variations in the detection
efficiency (\S \ref{sec:2E}) can induce spurious light curve effects
in the event that the image moves around in the focal plane.

A second caveat is formed by the inter-junction dead spaces (\S
\ref{sec:1}). These gaps cause, as the image moves over the array, a
time-dependent flux loss leading to a more-or-less random noise
component in the light curves. Simulations show that the photometric
scatter introduced by this effect is negligibly small ($\ll$1\%) for
typical seeing and image wander conditions; only in the rare case the
seeing is better (smaller) than the size of an individual pixel, i.e.,
$\sim$$0.60^{\prime\prime}$, the effect of the dead zones can be
significant (up to a few per cent). It is, in principle, possible to
partially correct for this effect in the data reduction process,
although the adequacy of the correction depends on the knowledge of
the instantaneous motion and form of the point spread function.

A third hazard is the image smearing component of atmospheric
seeing. Periods of poor seeing are characterized by a large point
spread function (full width at half maximum up to several arcseconds),
the extended wings of which may contain a significant fraction of the
total flux. Whereas such a configuration may already lead to a loss of
light beyond the detector boundary, the additional presence of image
motion can lead to significantly spurious light curve effects. These
effects are (potentially) energy-dependent as a result of atmospheric
dispersion (\S \ref{sec:4D}). An additional complication is posed by
the fact that poor seeing may, also depending on the brightness of the
object, place a significant number of source photons, compared to the
sky background, on the corner pixels $(X,Y = 1,1)$ and $(6,6)$ which
are used for sky subtraction (cf.\ Figure~\ref{fig:3}). Due to the
complexity of this process, an {\it a posteriori\/} correction for
these effects (similar to an ``aperture correction'' that is normally
applied in stellar photometry; e.g., Stetson 1990) is hardly
feasible. The effects of poor seeing can only be compensated for by
brute force, i.e., a larger focal-plane array.

\section{S--Cam data products}
\label{sec:4}

\subsection{Event-list data}
\label{sec:4A}

A key feature of S--Cam is its photon-counting ability. As a result,
the most rudimentary output of the pipeline processing is a
gain-corrected (and possibly barycentre-corrected) event
list. Although such data are potentially cumbersome to process and
interpret, knowledge of photon arrival times can be important, e.g.,
for high-precision ephemeris studies or timing of light pulses
(Delamere 1992; Perryman et al.\ 1999). Furthermore, the (partially)
processed event-list data allows the user to inspect and analyze the
observations at different time resolutions.

\subsection{Pixel- and energy-selected time-binned data}
\label{sec:4B}

Pixel- and energy-selected light curves are valuable for assessing the
performance of S--Cam2 and its data reduction software, in addition to
the astrophysical interpretation.

UZ For, the object discussed in this paper, is a cataclysmic variable:
a spatially unresolved eclipsing binary system composed of a white
dwarf and a low-mass main-sequence star filling its Roche lobe, in
which the highly magnetic degenerate object accretes material from its
mass-losing companion. During the 2~hour orbital period, the bright
bluish white dwarf is eclipsed for $\sim$8~minutes by the reddish
main-sequence star. This event is clearly seen in our data (e.g.,
Figure~\ref{fig:1}; cf.\ Reynolds 2001). Figure~\ref{fig:4} shows the
``flat-fielded'' and sky-subtracted pixel-integrated light curves of
UZ For in the energy bands red (``R''), yellow (``V''), and blue
(``B''; \S \ref{sec:2C}). (Although these bands were defined initially
such that they contain (roughly) equal amounts of counts (\S
\ref{sec:2C}), the subsequently applied atmospheric extinction
correction (\S \ref{sec:2F}), which is stronger in the blue than in
the red, has led to a significantly uneven distribution of counts in
the resulting pipeline product.) During the eclipse of the white
dwarf, the reddish light of the companion remains visible (cf.\
Figure~\ref{fig:3}). Figure~\ref{fig:4} also displays the hardness
ratio plots ``B/V'' and ``V/R'', which clearly show that the spectral
energy distribution of the observed light changes significantly during
(ingress and egress of) the eclipse (cf.\ \S \ref{sec:4C}). High
time-resolution spectrally-resolved data such as those presented here
provide strong constraints on, e.g., the geometry of the accretion
flow and the location of the accretion spots on the surface of the
white dwarf (e.g., Perryman et al.\ 2001).

\subsection{Spectra}
\label{sec:4C}

Figure~\ref{fig:5} shows the observed gain-corrected spectral energy
distribution of UZ For (plus sky) outside and during the eclipse of
the white dwarf (open squares and filled circles, respectively). It
clearly confirms the inherent energy resolution of S--Cam, another key
feature of the detector.

Extraction of intrinsic source spectra is a non-trivial exercise and
requires a detailed characterization of the transmission of all
components in the optical path, such as the atmosphere/night sky, the
WHT mirrors and derotator, and the S--Cam unit itself. STJ devices
nonetheless hold great promise as low-resolution spectro-photometers
for, e.g., extragalactic astrophysics. A carefully designed observing
and calibration program will ultimately allow for an efficient and
reliable determination of ``photometric redshifts'' of high-$z$
galaxies (e.g., Perryman et al.\ 1993; Jakobsen 1999).

\subsection{Image location: differential atmospheric refraction}
\label{sec:4D}

Atmospheric refraction causes a star at a zenith angle $z$ to move
towards the zenith by an amount $\sim$$(n - 1)~ \tan(z)$, where $n$ is
the index of refraction of the air layer near the Earth's surface. As
$n$ not only varies with atmospheric pressure $p$ and temperature $T$
but also with wavelength $\lambda$, refraction makes every object
appear as a little spectrum with the blue end towards the zenith and
the red end towards the horizon. This effect is known as atmospheric
dispersion or differential atmospheric refraction.

Atmospheric dispersion is clearly visible in the UZ For exposure (cf.\
Reynolds 2001): before the eclipse, i.e., during the first 10 minutes
of the exposure, the mean centroid locations of red and blue counts
(energy bands 1 and 3, respectively; \S \ref{sec:2C} and
Table~\ref{tab:1}) differ by $0.75^{\prime\prime}$. The amount of
atmospheric dispersion between energy bands 1 and 3 can also be
predicted by using, e.g., $n = n_\lambda(p,T)$ from Filippenko
(1982). Taking $p = 771$~mbar and $T = 14{\ }^{\circ}\! {\rm C}$ from
the night's observing log, we estimate that differential refraction
amounts to $\sim$$[n_{\lambda = 396~{\rm nm}} - n_{\lambda = 573~{\rm
nm}}]~ \tan(56^{\circ}\!) = 1.21^{\prime\prime}$, which is
significantly larger than the observed value. This prediction,
however, depends sensitively on the assumed wavelengths, among other
factors, and is best regarded as an order-of-magnitude estimate. The
prediction of the position angle of the direction in which atmospheric
dispersion operates on the S--Cam2 device, on the other hand, differs
by less than $3^{\circ}\!$ from the observed value, confirming the
hypothesis that the image centroid shift with wavelength is due to
atmospheric dispersion (cf.\ Figure~\ref{fig:6}).

\subsection{Image motion}
\label{sec:4E}

\begin{figure}
\centerline{\psfig{file=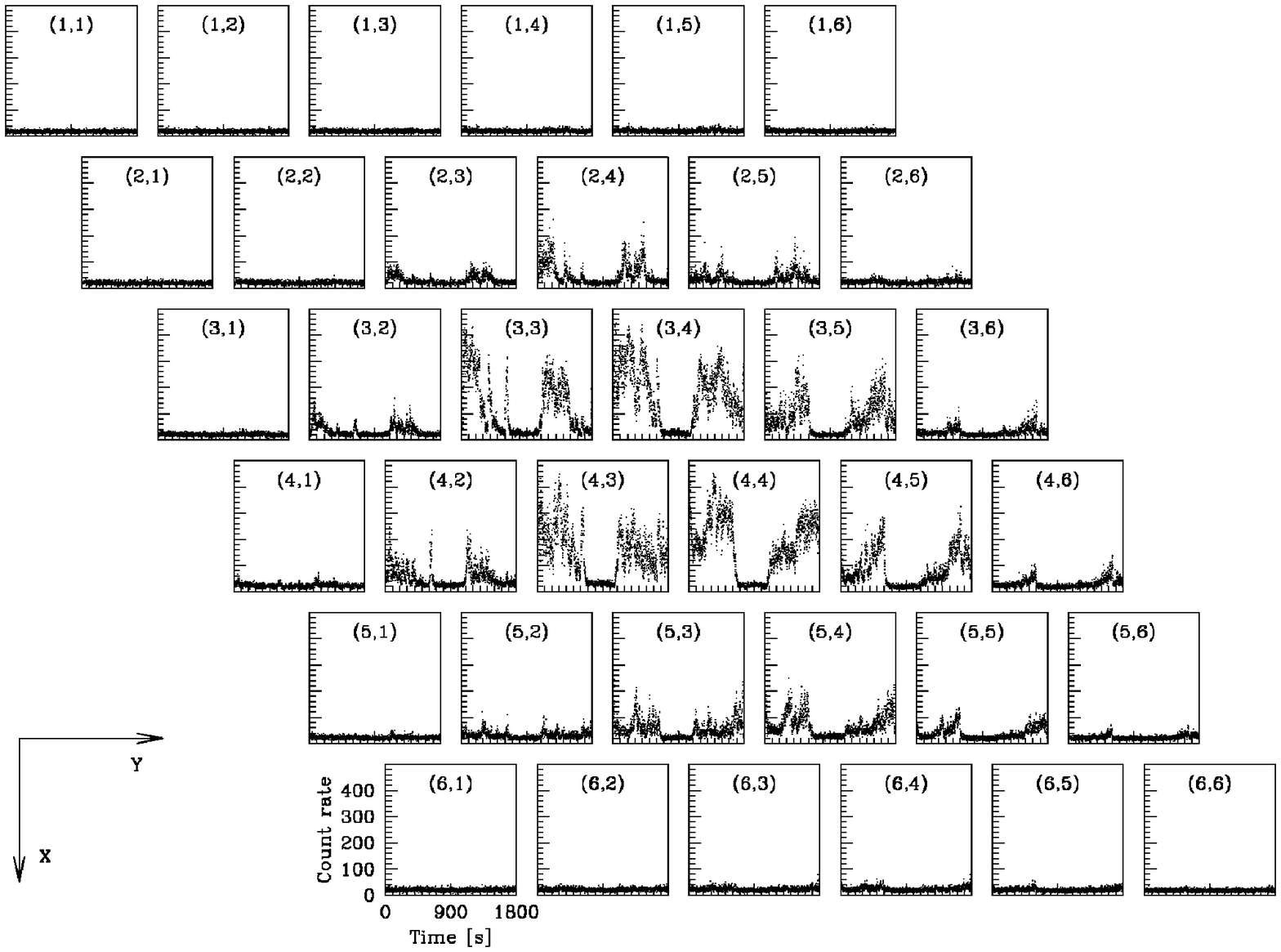,width=\textwidth,silent=,clip=}}
\caption{\null\hfill\break S--Cam2 array layout and pixel-selected
light curves of UZ For. The panels represent the genuine architecture
of the $6 \times 6$ array elements $(X,Y)$ in the focal plane. One
pixel is $25 \times 25~\mu$m$^2$ in size and spans
$\sim$$0.6^{\prime\prime} \times 0.6^{\prime\prime}$ on the
sky. Pixels are not directly adjacent: inter-junction dead spaces are
$\sim$$4~\mu$m. The data displayed in the pixels are light curves (at
a 1-s time resolution for all energies; \S\S
\ref{sec:2C}--\ref{sec:2D}) after gain correction (\S \ref{sec:2A}),
but before ``flat fielding'' (\S \ref{sec:2E}), extinction correction
(\S \ref{sec:2F}), and sky subtraction (\S \ref{sec:2G}; cf.\
Figure~\ref{fig:3}). Count rates are given in units of photons per
second; the time axis is relative to the start of the exposure (cf.\
\S \ref{sec:2B}). The pronounced dip in the light curves roughly
halfway the observation indicates an eclipse of the white dwarf in the
UZ For binary system (\S \ref{sec:4B}). The data clearly show that,
during the exposure, atmospheric seeing has moved the image around
over the array (\S \ref{sec:4E}). A number of sharp features, e.g.,
around ${\rm time} = 609$~s, corresponds to manual guiding corrections
(\S \ref{sec:4E}).}
\label{fig:1}
\end{figure}

\begin{figure}
\psfig{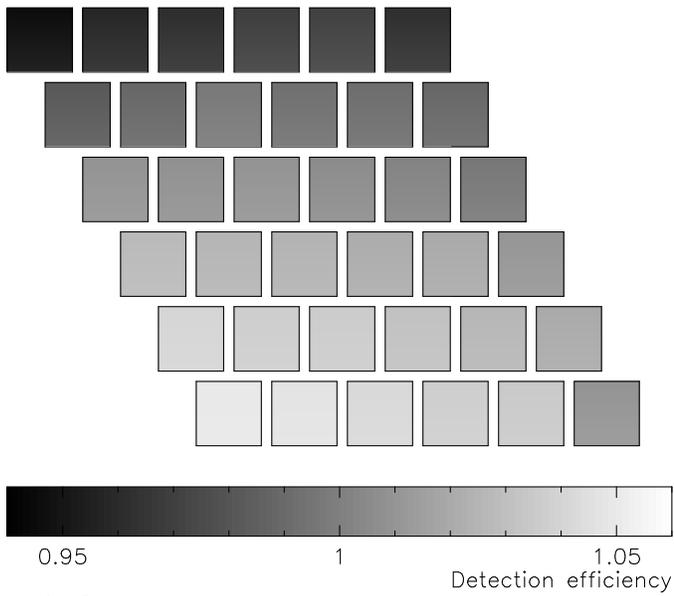}
\caption{\null\hfill\break Grey-scale representation of the detector
efficiency map. The responsivity of the reference pixel ($X,Y = 3,1$;
the first junction in the third row; cf.\ Figure~\ref{fig:1}) was
fixed at unity. The apparently smooth structure in the map is most
likely due to vignetting caused by the relay optics including the
infrared photon blocking filters (\S \ref{sec:2E}).}
\label{fig:2}
\end{figure}

\begin{figure}
\centerline{\psfig{file=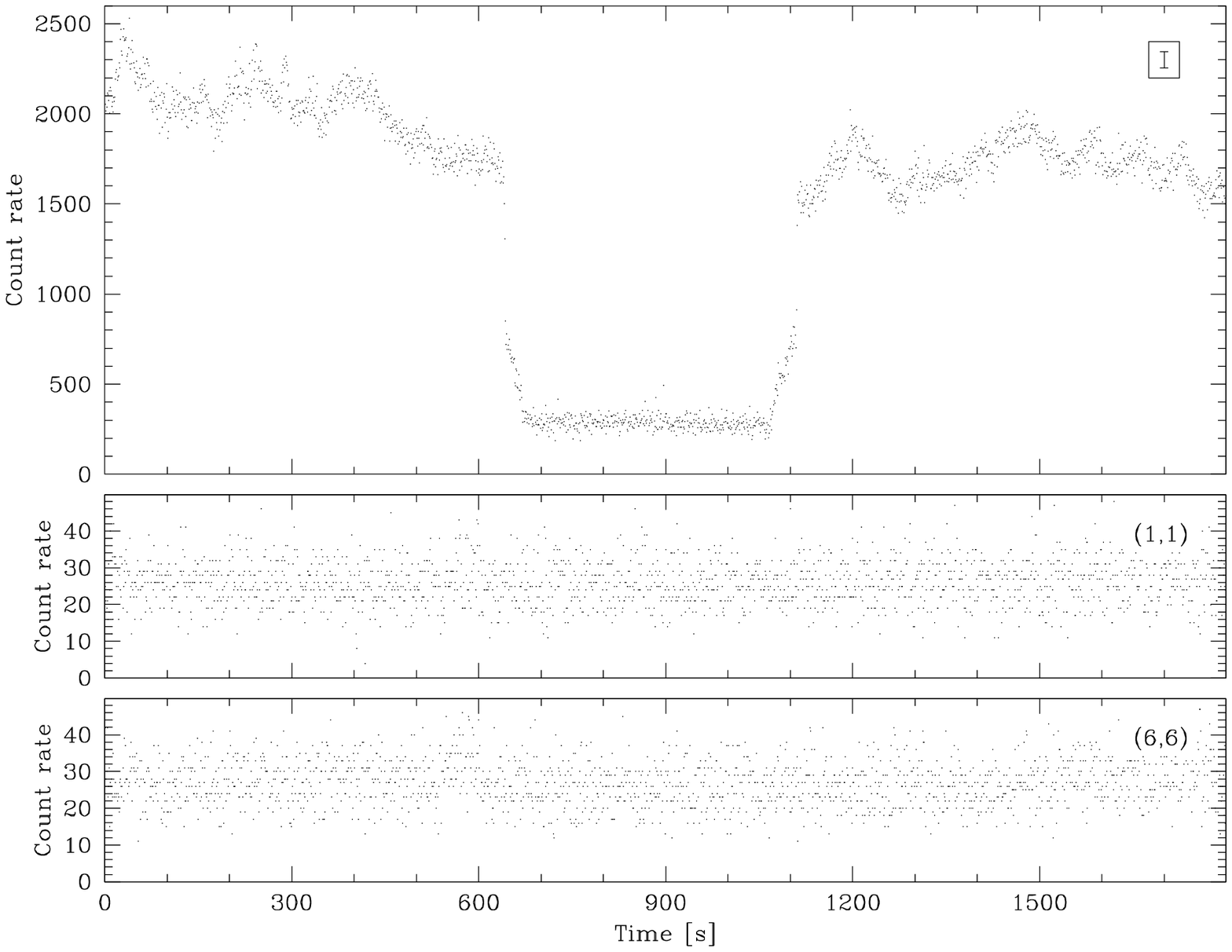,width=0.90\textwidth,silent=,clip=}}
\caption{\null\hfill\break Background subtraction. {\it Top panel:\/}
``flat-fielded'' and sky-subtracted pixel-integrated light curve for
UZ For (at a 1-s time resolution; count rate in photons per
second). The sky contribution was estimated at $6 \times 6 \cdot 26.5
= 954$~counts~s$^{-1}$, independent of time (see below; cf.\ \S
\ref{sec:2G}). The error bar in the top right corner roughly indicates
the expected Poisson noise for a count rate of
$2\,000$~counts~s$^{-1}$ (\S \ref{sec:3A}). During the eclipse of the
white dwarf, its faint companion accounts for
$\sim$$300$~photons~s$^{-1}$. The pixel-integrated light curve does
not reveal traces of the manual guiding corrections visible in the
pixel-selected data (Figure~\ref{fig:1}). {\it Bottom panels:\/} light
curves of the corner pixels $(X,Y = 1,1)$ and $(6,6)$ (count rates in
photons per second per pixel). S--Cam2 is a photon counting detector:
as a result, count rates take on integer values exclusively. The mean
time-averaged count rate of the two pixels is
$26.5$~photons~s$^{-1}$~pixel$^{-1}$; this value is assumed to
represent the sky background intensity. Around ${\rm time} \sim
550$~s, the image center slowly moves towards pixel $(6,6)$, as
evident from the slightly enhanced count rate (bottom panel); a manual
guiding correction at ${\rm time} = 609$~s corrects for this drift (\S
\ref{sec:4E} and Figure~\ref{fig:6}).}
\label{fig:3}
\end{figure}

\begin{figure}
\centerline{\psfig{file=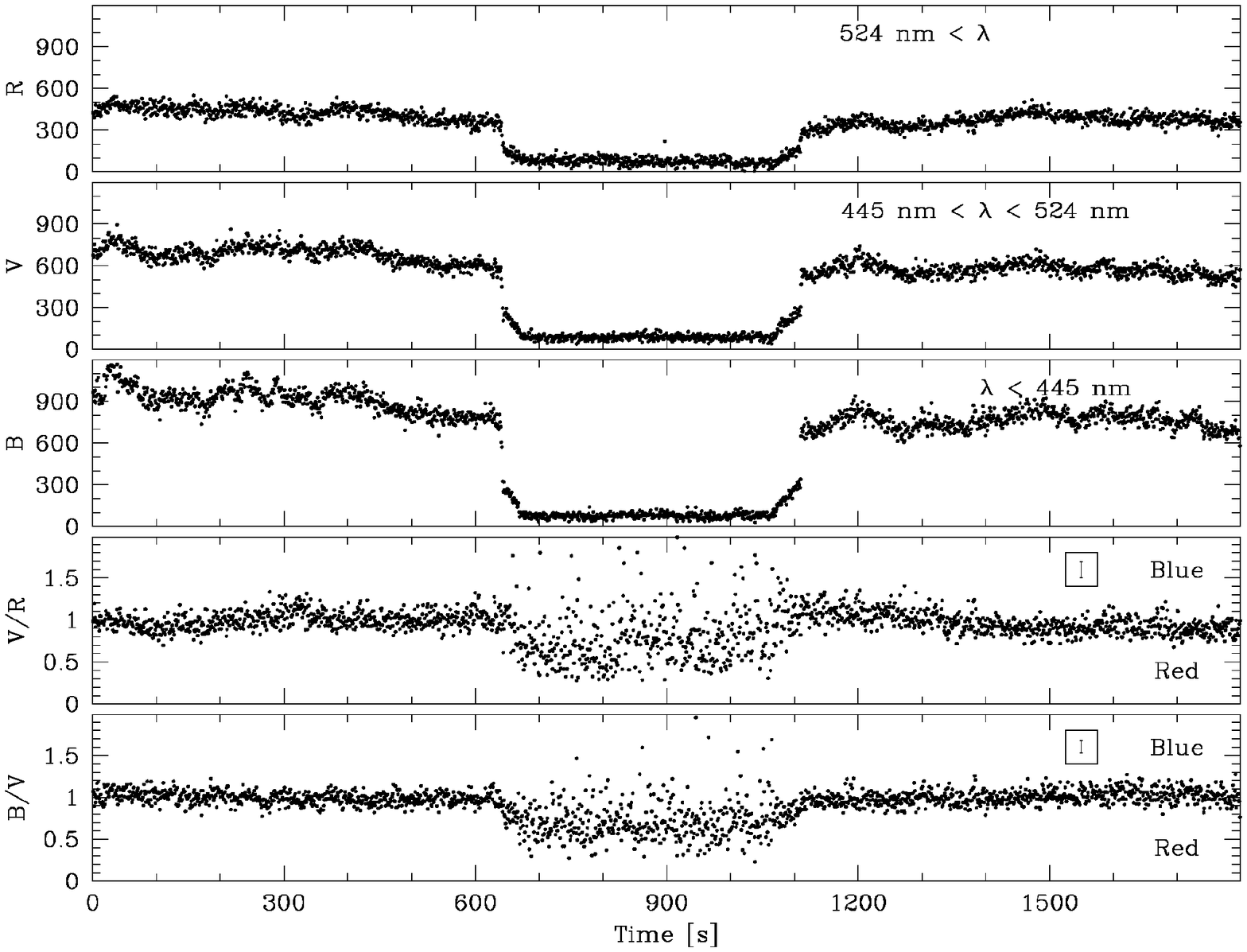,width=\textwidth,silent=,clip=}}
\caption{\null\hfill\break {\it Top three panels:\/} ``flat-fielded''
and sky-subtracted pixel-integrated light curves of UZ For in a low-,
medium-, and high-energy band, respectively. For ease of reference,
these bands are referred to as red (``R''; $E_{0}$--$E_{98}$; $\lambda
> 524$~nm), yellow (``V''; $E_{99}$--$E_{116}$; $445~{\rm nm} \leq
\lambda \leq 524$~nm), and blue (``B''; $E_{117}$--$E_{255}$; $\lambda
< 445$~nm; \S \ref{sec:2C} and Table~\ref{tab:1}). The energy bands
are defined such that they contain (roughly) equal amounts of counts;
later correction for atmospheric extinction (\S \ref{sec:2F}) leads to
a significantly uneven distribution. During the eclipse of the white
dwarf, the reddish light of the companion remains visible (cf.\
Figures~\ref{fig:3} and \ref{fig:5}). {\it Bottom two panels:\/} the
colour ratios ``V/R'' and ``B/V'' (after an arbitrary vertical
scaling); lower values indicate a redder colour. The error bars
indicate the expected noise due to photon statistics for artifical yet
representative count rates of $400, 600$, and $800$~counts~s$^{-1}$
for ``R'', ``V'', and ``B'', respectively. The spectral energy
distribution of the observed light changes significantly during the
eclipse of the white dwarf (cf.\ \S \ref{sec:4C} and
Figure~\ref{fig:5}).}
\label{fig:4}
\end{figure}

\begin{figure}
\psfig{file=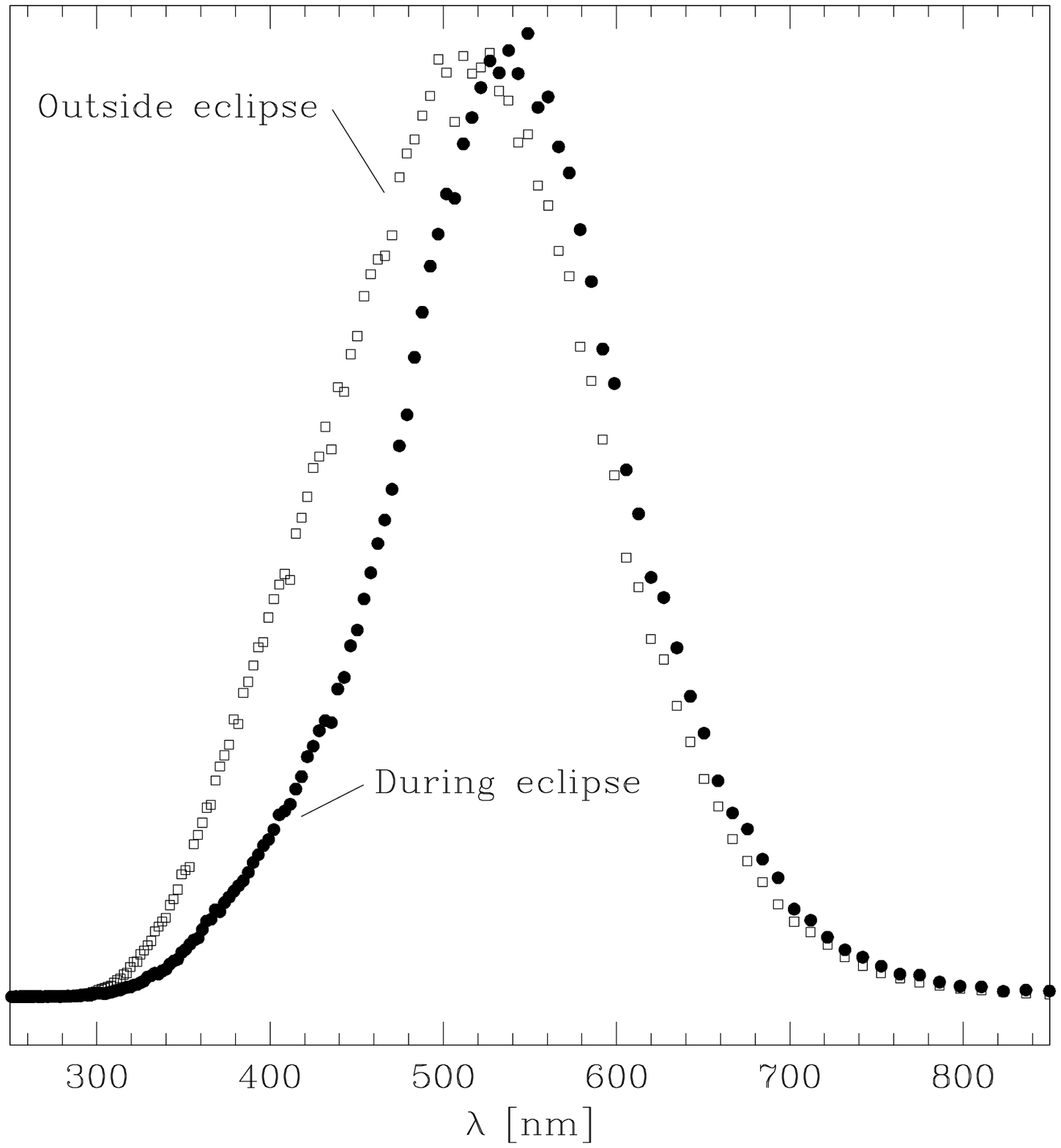,width=0.5\textwidth,silent=,clip=}
\caption{\null\hfill\break Gain-corrected and time-integrated spectral
energy distributions, before sky subtraction, during and outside the
eclipse of the white dwarf in UZ For (filled circles and open squares,
respectively). Wavelengths were obtained from energy channels by means
of equation~\ref{eq:3}. The vertical axis denotes counts per energy
channel on an arbitrary scale. The sky and faint red companion
contribute $\sim$$1\,000$ and $\sim$300~photons~s$^{-1}$ to the total
flux, respectively; the hot white dwarf accounts for
$\sim$$1\,500$~photons~s$^{-1}$ (e.g., Figure~\ref{fig:3}). The
spectrum outside eclipse is bluish, but with a significant
contribution from the sky; the spectrum during eclipse is dominated by
sky photons. The spectra nonetheless confirm the inherent energy
resolution of optical STJ detectors. ``Noisy'' features are due to the
analogue-to-digital conversion electronics and can be corrected during
data processing. The shape of the observed spectra can be understood
by convolving the intrinsic spectrum with the atmosphere, WHT mirrors
and derotator, S--Cam optics, detector efficiency, and detector energy
resolution. The blue cutoff at $\sim$310~nm is primarily due to the
atmosphere; the cutoff at $\sim$720~nm (taking the energy resolution
of S--Cam2 into account) is due to the infrared photon blocking
filters (\S \ref{sec:2E}).}
\label{fig:5}
\end{figure}

\begin{figure}
{\centerline{\psfig{file=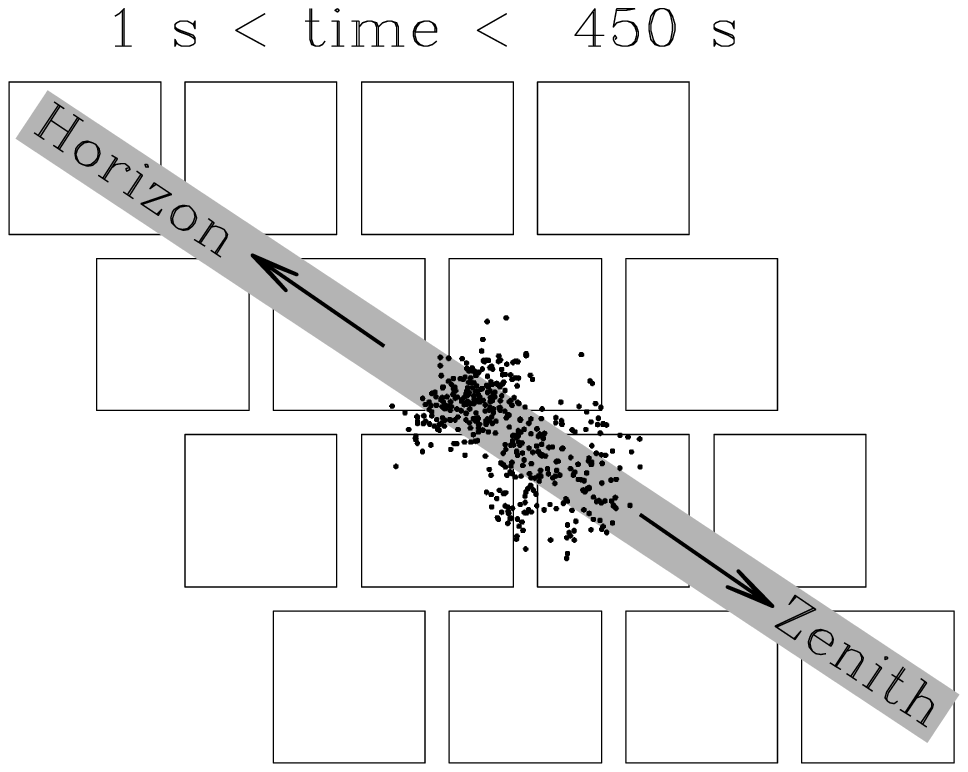,width=0.25\textwidth,silent=,clip=}\psfig{file=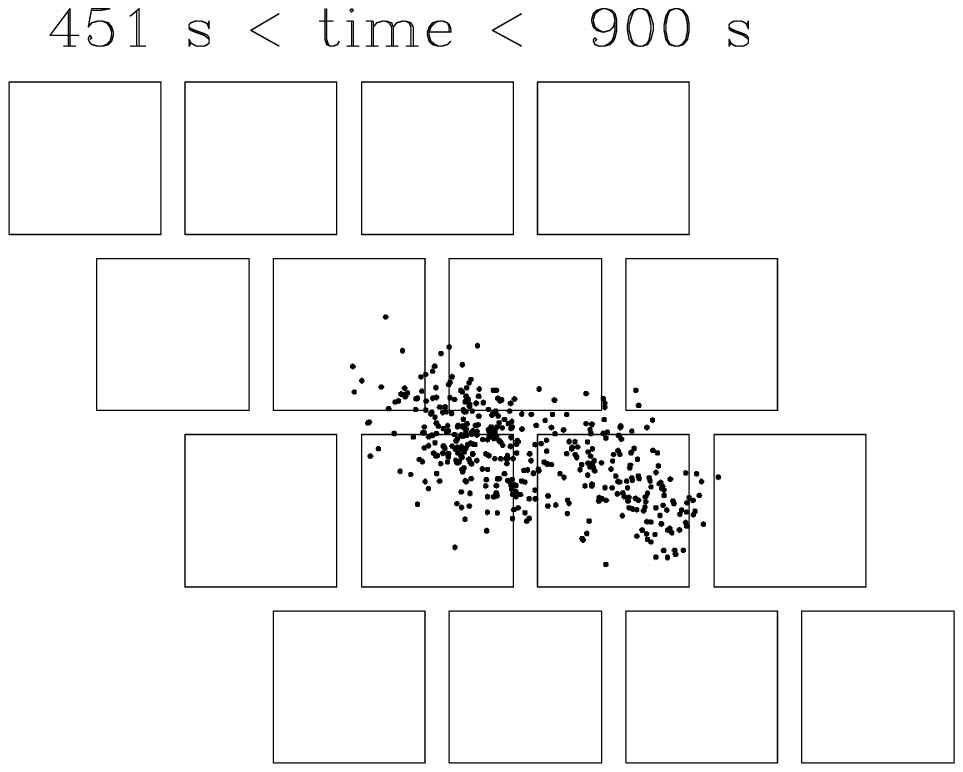,width=0.25\textwidth,silent=,clip=}\psfig{file=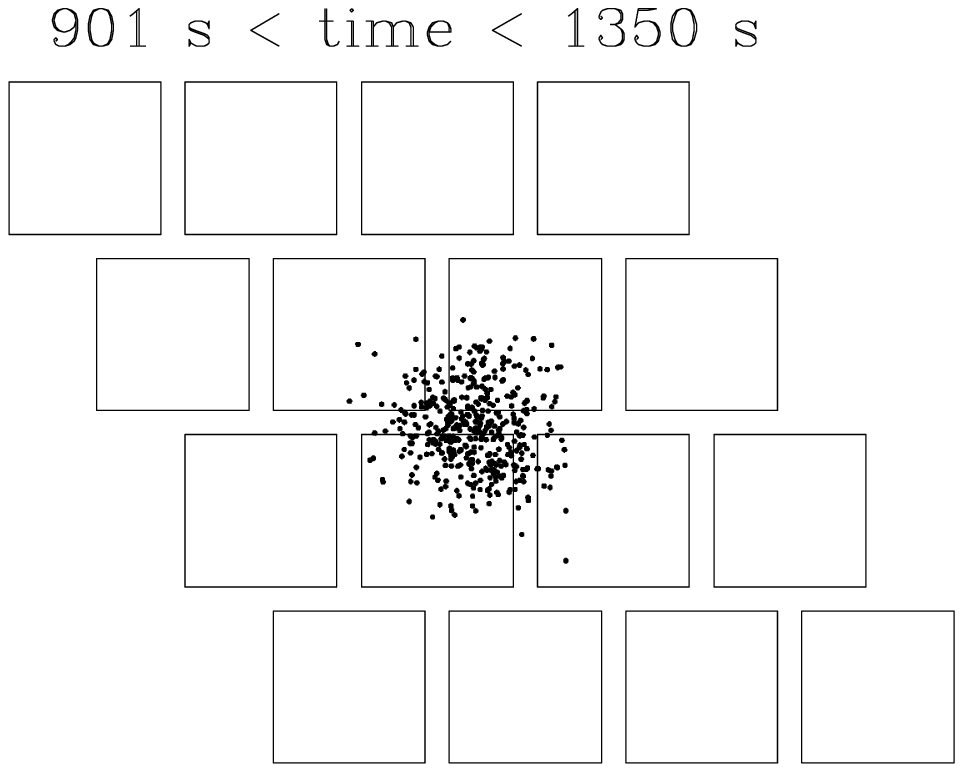,width=0.25\textwidth,silent=,clip=}\psfig{file=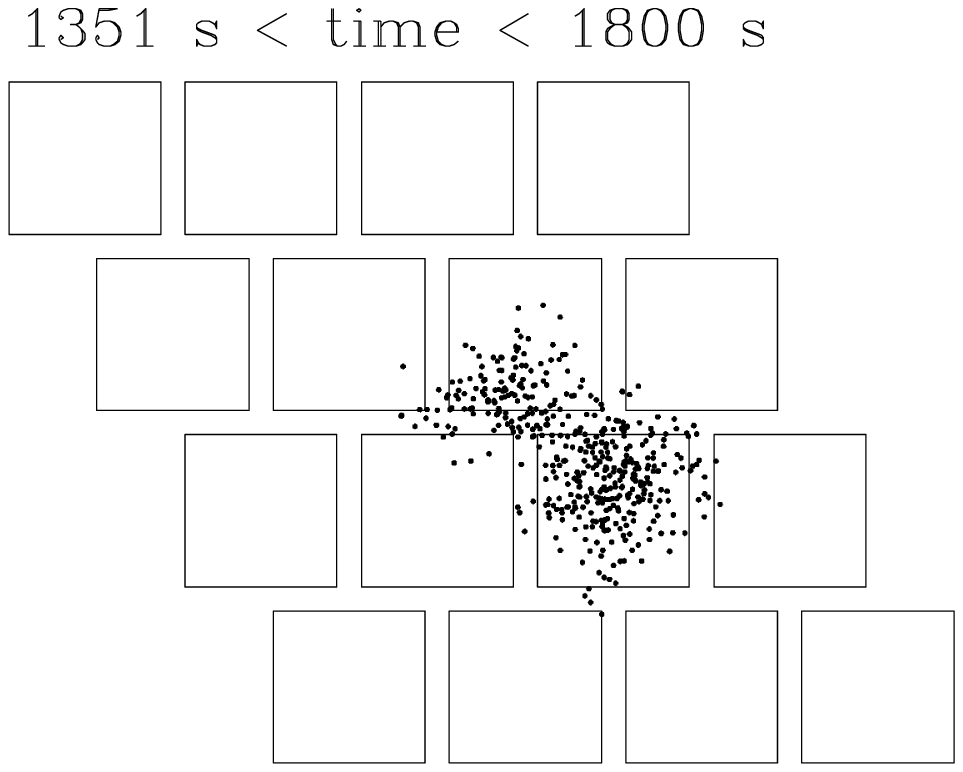,width=0.25\textwidth,silent=,clip=}}}
\caption{\null\hfill\break UZ For image centroid motion in the focal
plane during four equal-length time spans; only the 16 central pixels
are shown. Taking inter-junction dead spaces into account, one pixel
corresponds to $\sim$$0.7^{\prime\prime} \times 0.7^{\prime\prime}$ in
this figure. The 1-s time resolution data correspond to the
``flat-fielded'' and sky-subtracted counts (all energies). The left
panel indicates, for the UZ For observation discussed here, the
direction towards the zenith and the horizon. Outside eclipse (see
below), the observed mean centroid displacement between any pair of
consecutive 1-s time bins is $\sim$$0.17~{\rm pixel} \sim
0.12^{\prime\prime}$. The image motion can be understood as the
superposition of (i) random components due to atmospheric seeing
(root-mean-square centroid variation $\sim$$0.36^{\prime\prime}$) and
telescope-induced guiding and tracking errors and (ii) a number of
jumps due to manual guiding corrections (times $369, 609, 1\,010$, and
$1\,175$~s; \S \ref{sec:4E}; cf.\ Figures~\ref{fig:1} and
\ref{fig:7}). During the eclipse of the white dwarf ($640~{\rm s}
\mathrel{{\hbox to 0pt{\lower
3pt\hbox{$\sim$}\hss}}\raise2.0pt\hbox{$<$}} {\rm time}
\mathrel{{\hbox to 0pt{\lower
3pt\hbox{$\sim$}\hss}}\raise2.0pt\hbox{$<$}} 1120 $~s), the count rate
of the faint companion ($\sim$300~photons~s$^{-1}$) is too low to
allow for a reliable determination of the centroid location at the
selected binning period (cf.\ \S \ref{sec:4F}).}
\label{fig:6}
\end{figure}

Figure~\ref{fig:6} shows the evolution of the centroid position of the
``flat-fielded'' and sky-subtracted counts (all energies; cf.\
Figure~\ref{fig:1}). An inspection of the data reveals that the image
centroid is relatively stable, except for a number of discontinuous
jumps. These features could, {\it a posteriori\/}, be related to (at
least) four small manual guiding corrections at times $369, 609,
1\,010$, and $1\,175$~s (cf.\ Figures~\ref{fig:1} and
\ref{fig:7}). Another feature in the image centroid location is
(potentially) caused by differential refraction: during the eclipse,
the source light becomes significantly less blue (e.g.,
Figures~\ref{fig:4}--\ref{fig:5}), with the result that the image
centroid effectively moves in the direction towards the horizon (but
see \S \ref{sec:4F}; cf.\ \S \ref{sec:4D}).

The image motion displayed in Figure~\ref{fig:6}, disregarding the
artificial discontinuities mentioned above, is partly explained by
telescope-induced tracking and guiding errors (cf.\ \S \ref{sec:4G}
and Wilson et al.\ 1999). The main contribution, however, is due to
atmospheric seeing. The expected root-mean-square image centroid
variation due to seeing is (e.g., Lindegren 1979):
\begin{equation}
\sim\ \sqrt{0.2} ~ \left({{\lambda}\over{D}}\right)^{1/6} \cdot \left({{\lambda}\over{D_{\rm eff}}}\right)^{5/6}~ \sim\ 0.36^{\prime\prime},
\label{eq:6}
\end{equation}
where $D=4.2$~m is the diameter of the aperture, $\lambda = 479$~nm is
the centroid wavelength (Table~\ref{tab:1}), and $D_{\rm eff} = 0.98~
\lambda / \epsilon = 0.06$~m is Fried's (1966) parameter; $\epsilon$
is the full width at half maximum of the atmospheric seeing-induced
point spread function ($\sim$$1.60^{\prime\prime}$ in this case; \S
\ref{sec:4F}). During the exposure, the image centroid is thus
expected to remain within $\sim$1 pixel, consistent with the
observations.

\subsection{Image size: atmospheric seeing}
\label{sec:4F}

The spatial resolution of S--Cam2 allows for the measurement not only
of the centroid of the point spread function (PSF) in the focal plane
(\S\S \ref{sec:4D} and \ref{sec:4E}) but also of its size. As the
combined effect of dome, mirror, and telescope seeing (due to
tracking, focus, and aberration errors) is generally small for the WHT
($\mathrel{{\hbox to 0pt{\lower
3pt\hbox{$\sim$}\hss}}\raise2.0pt\hbox{$<$}}$$0.30^{\prime\prime}$;
e.g., Mu\~noz--Tu\~n\'on et al.\ 1998; Wilson et al.\ 1999), S--Cam2
can be used as an atmospheric seeing monitor. Seeing is usually
expressed in terms of the full width at half maximum (FWHM, in units
of arcseconds) of the time-averaged PSF (e.g., Coulman 1985). As the
typical time scale of seeing is on the order of a few ms, a 1-s time
binning generally allows for a sufficiently well-defined (i.e.,
smooth) PSF.

\begin{figure}
\centerline{\psfig{file=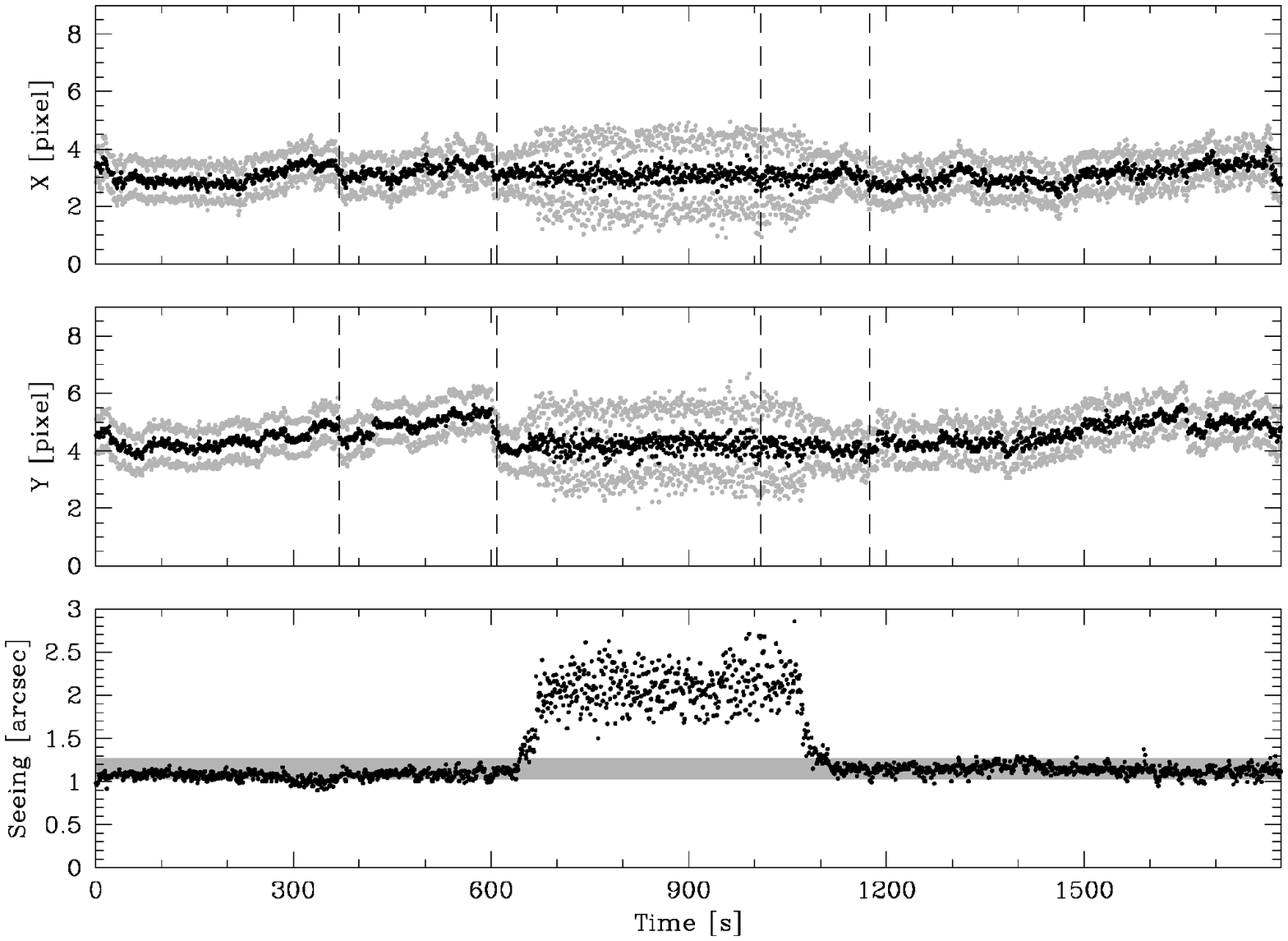,width=\textwidth,silent=,clip=}}
\caption{\null\hfill\break Atmospheric seeing analysis for UZ
For. {\it Top panel:\/} the evolution of the centroid position (black
dots) and the corresponding interquartile range (grey dots) in the $X$
direction (the ``vertical'' direction on the device; cf.\
Figure~\ref{fig:1}). The data, at a 1-s time resolution, refer to the
``flat-fielded'' and sky-subtracted counts (all energies). During the
eclipse of the white dwarf, the number of source counts from the faint
companion is insufficient to determine the centroid and interquartile
range reliably at the selected time resolution (\S
\ref{sec:4F}). Manual guiding corrections were made around times $369,
609, 1\,010$, and $1\,175$~s; these explain discontinuities in the
centroid evolution (cf.\ Figure~\ref{fig:6}). {\it Second panel:\/} as
the top panel, but for the ``horizontal'' $Y$ direction on the
device. {\it Bottom panel:\/} inferred seeing estimates with time
(corrected for zenith angle) assuming a Gaussian point spread function
(\S \ref{sec:4F}). The grey band denotes the mean atmospheric seeing
estimate during the last 700~s of the exposure: $\epsilon =
1.14^{\prime\prime}$ ($1.60^{\prime\prime}$ uncorrected for zenith
angle). The increased seeing during the eclipse is spurious (\S
\ref{sec:4F}).}
\label{fig:7}
\end{figure}

The top panels of Figure~\ref{fig:7} show, for the ``vertical'' ($X$)
and ``horizontal'' ($Y$) directions, the evolution of the centroid
position (black dots; \S \ref{sec:4E}) and the interquartile range
(grey dots) of the ``flat-fielded'' and sky-subtracted data (all
energies). From these curves, the size of the region in which half of
all counts is contained is readily obtained. If the functional form of
the PSF is known (or specified), the size of this ``interquartile
area'' can be directly related to the FWHM $\epsilon$, i.e., the
amount of atmospheric seeing. The final step transforms $\epsilon$ to
units of arcseconds and corrects for zenith angle by multiplying with
${\cos}^{3/5}(z)$. The bottom panel of Figure~\ref{fig:7} shows the
evolution of the inferred seeing during the UZ For exposure assuming a
Gaussian PSF; a Moffat (1969) function generally returns smaller
estimates. The mean atmospheric seeing after the eclipse (see below)
is $1.60^{\prime\prime}$ ($1.14^{\prime\prime}$ corrected for zenith
angle). These estimates seem robust and sensible; unfortunately, an
independent measurement of the seeing at the time of the observation
is unavailable.

An uncertainty in the procedure outlined above is introduced by the
use of polychromatic light, which is susceptible to differential
atmospheric refraction (\S \ref{sec:4D}), for the definition of the
interquartile area. In the case of UZ For, atmospheric dispersion
operates, by coincidence, nearly along the diagonal direction on the
device (e.g., Figure~\ref{fig:6}), so that the PSF is elongated
roughly equally strongly in the $X$ and $Y$ directions. The net result
of this effect therefore is an overestimate of the size of the
interquartile area, i.e., an overestimate of the inferred seeing, by
$\sim$$10$\%. The latter number was obtained by repeating the seeing
analysis outlined above twice while restricting ourselves to
semi-monochromatic light, i.e., ``blue'' and ``red'' counts.  These
tests also indicate that the mean interquartile ranges in $X$ and $Y$
for semi-monochromatic light may differ by $\sim$$10$\% from each
other, i.e., even the semi-monochromatic PSF is slightly asymmetric.

A crucial assumption in the procedure described above is that, given
the selected time resolution, a sufficient number of source counts is
available for a reliable definition of the image centroid and
interquartile area. During the eclipse of the white dwarf by its faint
companion this assumption is clearly violated at the selected time
resolution (1~s; cf.\ Figure~\ref{fig:3}). This is evident, e.g., from
the fact that the interquartile area seems significantly larger during
than outside eclipse. We therefore conclude that the inferred seeing
increase during the eclipse of the white dwarf may be spurious.

\subsection{Telescope pointing}
\label{sec:4G}

The high speed and photon-counting and imaging capabilities of S--Cam2
enable the user to assess the impact on the final light curves of
issues related to telescope tracking and guiding. The WHT has an
altitude--azimuth mounting with two independent motors; its autoguider
system operates at $\sim$$0.1$~Hz, i.e., tracking errors occurring at
higher frequencies are uncorrected for.

\begin{figure}
\centerline{\psfig{file=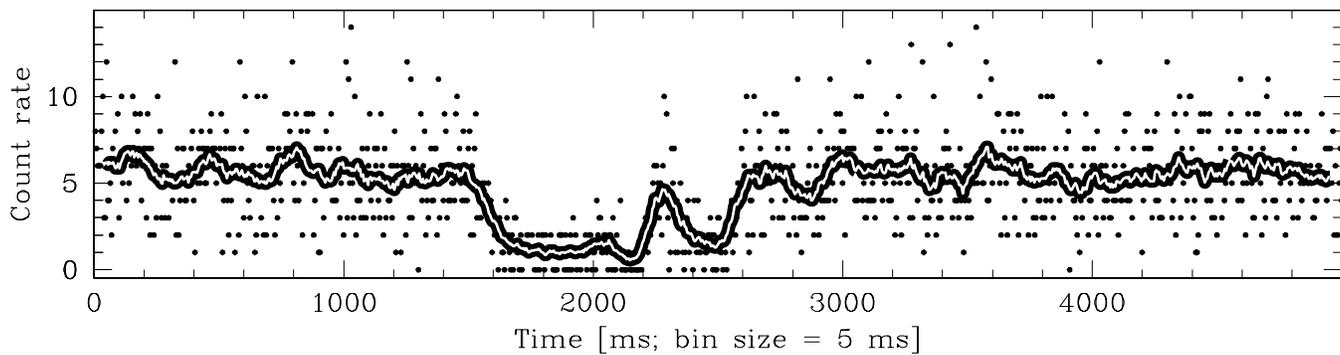,width=\textwidth,silent=,clip=}}
\caption{\null\hfill\break Pointing glitch in the UZ For exposure (\S
\ref{sec:4G}). The dots show the light curve of the source (all
energies), before extinction correction and background subtraction,
over a 5-s period at a time resolution of 5~ms; the ordinate is in
units of photons per time bin. The curve is the running average of the
count rate over 20 time bins, i.e., 0.1~s. Glitch ``ingress'' runs
from time $\sim$$1\,500$ to $\sim$$1\,700$. During $\sim$$0.3$~s
following ``ingress'', S--Cam2 registers $\sim$0--2 sky photons per
5~ms time bin, compared to $\sim$$5$~counts per time bin
normally. Glitch ``egress'' runs from time $\sim$$2\,000$ to
$\sim$$3\,000$. The spatial resolution of S--Cam2 reveals that the
dis- and re-appearance of source light takes place along the azimuthal
direction (not shown).}
\label{fig:8}
\end{figure}

\noindent{\bf Pointing glitches:} The azimuth drive is known to suffer
from pointing glitches (see {\tt
http://www.ing.iac.es/$^\sim$crb/wht/tracking.html}). During these
unpredictable events, the telescope pointing shoots off in azimuth
after which it returns to its nominal position. The typical spatial
amplitude of such glitches is several arcseconds; the typical duration
is $\sim$1.5~s. An S--Cam2 observation of UZ For obtained during the
Dec 1999 campaign, among others, contains two of these events (cf.\
Perryman et al.\ 2001). Figure~\ref{fig:8} shows the pixel-integrated
light curve of the source over a 5-s period spanning the first glitch,
at a resolution of 5~ms; a higher time resolution is precluded by the
low photon flux of UZ For.

Glitch ``ingress'' takes $\sim$$0.2$~s (${\rm time}$
$\sim$$1\,500$--$1\,700$), during which the image centroid is seen
shooting off the device in the azimuthal direction (not shown). It
takes $\sim$$0.3$~s for the WHT system to react to the loss of light
(${\rm time}$ $\sim$$1\,700$--$2\,000$). In correcting for the glitch,
the telescope initially ``overshoots'' (time $\sim$2\,300), after
which it ``bounces back'' along the azimuthal direction (not shown) to
its nominal position in $\sim$$1.0$~s (${\rm time}$
$\sim$$2\,000$--$3\,000$). The full duration of this glitch is
$\sim$$1.5$~s. An S--Cam2 analysis of several of these features has
revealed that their characteristics are very similar. It is not
exactly known what triggers these pointing anomalies; large zenith
angles could play a role ($z \sim 58.4^{\circ}\!$ in this case).

\begin{figure}
\psfig{file=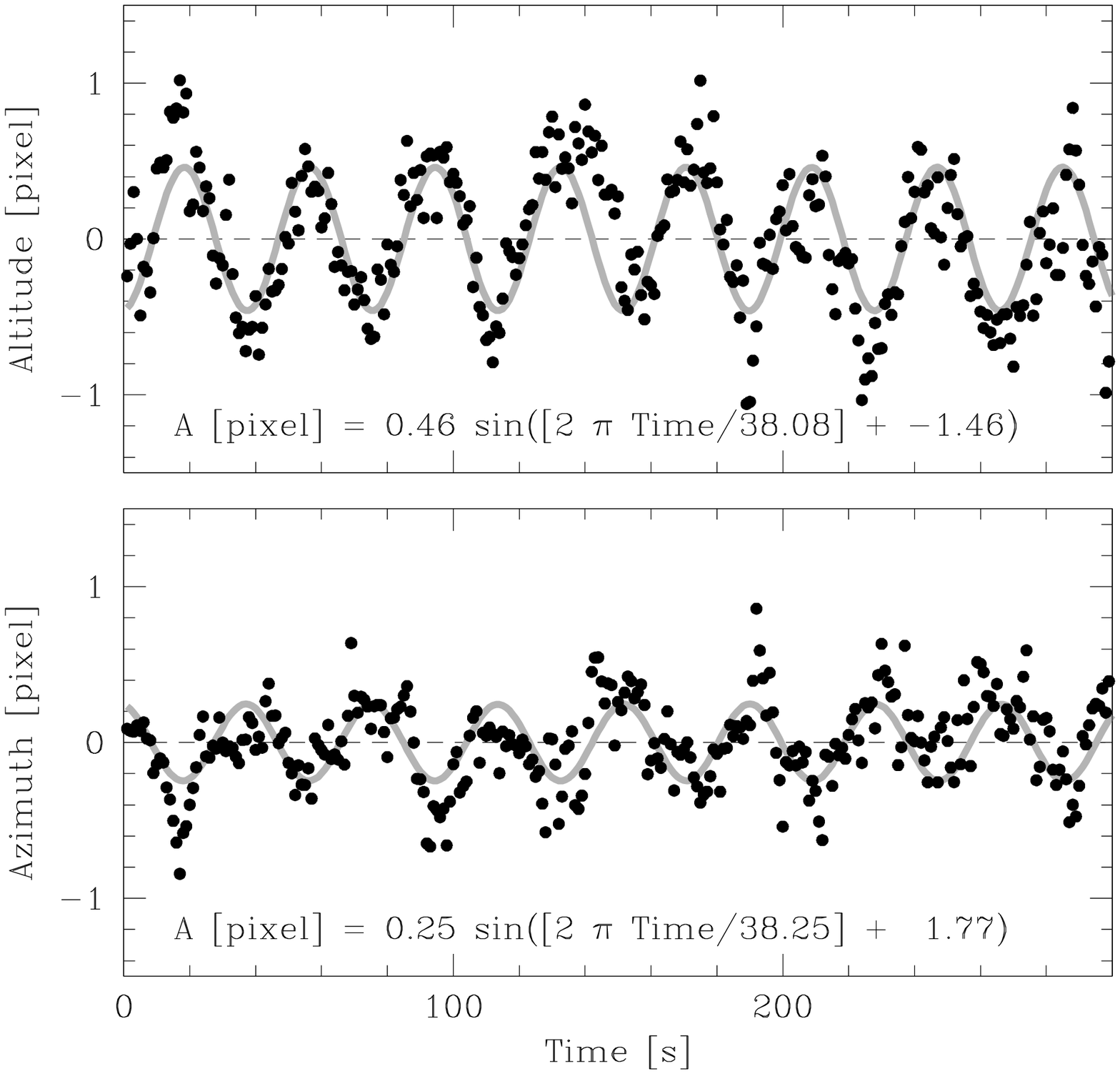,width=0.5\textwidth,silent=,clip=}
\caption{\null\hfill\break Pointing oscillation (\S \ref{sec:4G}). The
dots show the evolution of the image centroid location with time, at a
1-s resolution, for the altitude (elevation; {\it top panel\/}) and
azimuth directions ({\it bottom panel\/}). The ordinates denote
location in the focal plane in units of pixels (one pixel is
$\sim$$0.7^{\prime\prime}$, after correcting for inter-pixel dead
spaces). The grey sine-curves are two independent fits to the data;
the best-fit parameters are indicated in the panels (amplitudes in
units of pixels, time and periods in units of seconds, and phases in
radians). The fits provide a fair representation of the observations,
given the facts (i) that the chosen functional form does not
necessarily have a physical meaning, and (ii) that a random image
motion component due to seeing is also present in the data (\S
\ref{sec:4E}).}
\label{fig:9}
\end{figure}

\noindent{\bf Pointing oscillations:} A remarkable finding was
obtained during the Sep--Oct 2000 campaign when observing a
non-variable circumpolar reference source for $300$~s. During this
exposure, the image centroid location showed an oscillatory pattern in
the focal plane (Figure~\ref{fig:9}), unlikely to have been caused by
random atmospheric phase fluctuations, i.e., seeing (\S
\ref{sec:4E}). The grey curves in Figure~\ref{fig:9} show two
independent sinusoidal fits to the image centroid data in the two
``principal directions'' altitude (elevation) and azimuth. Although
this functional form is not necessarily the optimum choice, and has
been chosen for illustrative purposes only, the fits provide a fair
representation of the data. They suggest that the pointing oscillation
has a period of $\sim$$38.1$~s ($\sim$$0.026$~Hz) and an amplitude of
$\sim$$0.36^{\prime\prime}$, 90\% of which is in the altitude
direction. The combination of the long period and relatively small
amplitude of this phenomenon (as compared to a typical seeing disk
with full width at half maximum of $\sim$$1^{\prime\prime}$) explains
why similar oscillations have not been reported previously.

The large zenith angle of the source ($z \sim 56.5^{\circ}\!$) might
be a key factor in understanding the cause of the pointing
oscillation. It is also interesting, in this respect, to recall that
the WHT has a known structural oscillation frequency at
$\sim$$2.7$~Hz. This resonance can have a significant effect if it is
strongly excited, e.g., by wind buffeting of the telescope structure
(see {\tt
http://www.ing.iac.es/Astronomy/development/hap/tracking.html}).
During the observation discussed here, wind speed recordings were
normal ($\sim$$30$--$35~{\rm km}~{\rm h}^{-1}$); the dome slit,
however, was pointed to within $\sim$$30^{\circ}\!$ of the direction
from which the wind was coming.

\section{Discussion and conclusions}
\label{sec:5}

Even as recently as a decade ago, a high-speed optical photon counting
detector with inherent energy resolution was an unknown concept. The
development of the first generation of such an instrument, based on
STJ technology, has now advanced to the stage where STJ-based imaging
cameras containing dozens of pixels are starting to be operated. The
high quality and potential of STJ data has been demonstrated, and
first astrophysical results based on such measurements have been
published (e.g., Perryman et al.\ 2001). With improved wavelength
resolution in the future, STJ detectors, being operated as
``multi-object redshift machines'', could significantly impact the
field of extragalactic astronomy. Earlier and ongoing investigations
aimed at evaluating an STJ-based instrument for the next-generation
space telescope (NGST; e.g., Jakobsen 1999), as an interferometric
fringe sensor for VLTI (R.S.~Le Poole and W.~Jaffe, private
communication), and as a polarimeter for Subaru (M.~Cropper, private
communication) are merely the beginning of the integration of this
novel detector into the forefront of astronomical thinking.

A number of STJ detectors in an array format, an STJ camera (S--Cam),
generates four-dimensional data: photon arrival time, energy, and
array element $(X,Y)$. The reduction and analysis of such data,
typically several thousands of photons per second, is potentially
involved and, while part of the high-energy data analysis culture,
differs significantly from procedures adopted in a CCD
environment. This paper describes an automated data reduction pipeline
that is aimed at processing data generated by ESA's S--Cam2
device. The implementation of the software is highly transparent and
guarantees its portability to different platforms as well as an easy
adaptation to future generation cameras. The pipeline takes care of
energy calibration, barycentric time stamp correction, energy range
selection, time binning, detector efficiency correction, atmospheric
extinction correction, and sky background subtraction; a future
version will also correct light-curve data for ``noise'' induced by
the dead spaces between the array elements.

The analysis and interpretation of STJ data differs, in some respects,
significantly from procedures adopted in CCD-based work. Differential
atmospheric refraction, for example, is clearly observable with an STJ
detector array and should be reckoned with. Atmosphere- or
telescope-induced image motion on the device can be responsible,
through a number of mechanisms, for spurious effects in light curves
and hardness ratio plots. The interpretation of such data thus
requires care. However, the sensitivity of S--Cam to such atmospheric
effects offers, at the same time, the possibility to deepen our
knowledge of the characteristics of atmospheric seeing, such as the
relative importance of the wings of the seeing profile, the statistics
of the amount of image smearing and motion, and wavelength-dependent
effects. S--Cam data also provide useful diagnostics in relation to
telescope pointing, tracking, and guiding.

Although revolutionary, the current generation of Tantalum-based STJ
cameras is amenable to further development. Optical-STJ research
within ESA is presently mainly driven towards the use of smaller
energy gap/lower critical temperature (i.e., higher energy resolution)
materials such as Hafnium, and towards larger format arrays (e.g.,
Verhoeve et al.\ 2001). Despite the fact that considerable technical
challenges are to be faced, it is envisaged that a Hafnium-based
detector array will ultimately allow for a wavelength resolution of
5~nm at 500~nm and count rates up to $10\,000$~photons per pixel. A
space-borne version of such an instrument, with a response from 100 nm
to 2~$\mu$m and a detection efficiency larger than 80\% throughout
this range, would play a major role in future astronomy.

\acknowledgments The research described in this paper is based on
observations made with the William Herschel Telescope operated on the
island of La Palma by the Isaac Newton Group (ING) in the Spanish
Observatorio del Roque de los Muchachos of the Instituto de
Astrof\'{\i}sica de Canarias. In ESTEC, N.~Rando, J.~Verveer,
S.~Andersson, D.~Martin, and P.~Verhoeve are thanked for their efforts
in developing the S--Cam instrument. We are grateful to the ING staff
for continued assistance, to the WHT time allocation committee for the
generous assignment of engineering time for the first commissioning of
S--Cam, and to our scientific collaborators involved in the analysis
of the UZ For data, M.~Cropper and G.~Ramsay (Mullard Space Science
Laboratory, University College London).

\vfill\eject

\vfill\eject

\vfill\eject

\vfill\eject

\vfill\eject

\vfill\eject

\vfill\eject

\vfill\eject

\vfill\eject

\begin{table}
\caption{\null\hfill\break Energy range selection and atmospheric
extinction correction for UZ For (\S\S \ref{sec:2C} and
\ref{sec:4D}). The first column denotes the energy band (0 for all
energies); the second column schematically denotes its colour. The
third and fifth columns denote the lower and upper energy channels
defining the energy band; the fourth and sixth columns denote its
centroid channel and wavelength, respectively
(equation~\ref{eq:3}). The last column provides the corresponding mean
extinction coefficient $k_\lambda$ valid for La Palma. The value for
the exposure containing all counts (energy band 0) is the mean of the
coefficients of the energy-selected sub-exposures; $k_\lambda$
corresponding to the centroid wavelength ($\lambda = 479$~nm) is
$0.1434$~mag~airmass$^{-1}$. The count rate correction factors are
$10^{0.4 k_\lambda \cdot A}$; the airmass $A$ changes from $1.75$ to
$1.83$ during this observation.}
\begin{tabular}{ccrrrcc}
Energy & ``colour''& $E_{\rm low}$ & $E_{\rm centroid}$ & $E_{\rm high}$ & $\lambda_{\rm centroid}$ & $k_{\lambda}$\\
band & & & & & nm & mag~airmass$^{-1}$\\
\hline
0 & white  &   0 & 108 & 255 & 479 & 0.1798\\
1 & red    &   0 &  90 &  98 & 573 & 0.0993\\
2 & yellow &  99 & 106 & 116 & 488 & 0.1345\\
3 & blue   & 117 & 131 & 255 & 396 & 0.3056
\end{tabular}
\label{tab:1}
\end{table}

\end{document}